\documentclass[twocolumn]{article}
\usepackage[margin=0.5in]{geometry}
\usepackage{amssymb}
\usepackage{graphicx} 
\usepackage{xcolor}
\usepackage[english]{babel} 
\usepackage{amsmath}
\usepackage{caption}
\usepackage{gensymb}
\usepackage{natbib}

\usepackage{bbm} % for the 1 symbol

\title{Plasmonic metamaterial time crystal}
\author{Tingwen Guo$^1$, Jules Sueiro$^2$, Gian Marcello Andolina$^2$, Artem Levchuk$^1$, Stefano Ponzoni$^1$,\\Romain Grasset$^1$, Donald Monthe$^1$, Ian Aupiais$^1$, Dmitri Daineka$^3$, Javier Briatico$^4$, Thales VAG de Oliveira$^5$, \\
Alexey Ponomaryov$^5$, Atiqa Arshad$^5$, Arjun Karimbana-Kandy$^5$, Gulloo Lal Prajapati$^5$, Igor Ilyakov$^5$, \\ Jan-Christoph Deinert$^5$, Sebastian F. Maehrlein $^{5,6}$, Luca Perfetti$^1$, Marco Schir\`o$^2$, Yannis Laplace$^1$}
\date{$^1$ Laboratoire des Solides Irradi\'es, CEA/DRF/IRAMIS, Ecole Polytechnique,
CNRS, Institut Polytechnique de Paris, F-91128 Palaiseau, France \\$^2$ JEIP, UAR 3573 CNRS, Collège de France, PSL Research University, 11 Place Marcelin Berthelot, F-75321 Paris, France \\$^3$  LPICM, CNRS, Ecole Polytechnique, Institut Polytechnique de Paris, F-91128 Palaiseau, France \\$^4$ Laboratoire Albert Fert, CNRS, Thales, Universit\'{e} Paris Saclay, 91767 Palaiseau, France \\
$^5$ Institute of Radiation Physics, Helmholtz-Zentrum Dresden-Rossendorf, Dresden, 01328, Germany \\
$^6$ Institute of Applied Physics, Dresden University of Technology, Dresden, 01062, Germany \\
} 

\begin{document}

\maketitle % Afficher le titre et les auteurs

\begin{abstract}

Periodically driven optical materials and metamaterials have recently emerged as a promising platform for realizing photonic time crystals (PTCs), which are systems whose optical properties are strongly and periodically modulated on timescales comparable to the optical cycle of light \cite{Asgari_2024}. These time-varying structures are the temporal counterparts of spatial photonic crystals (SPCs), for which a large and periodic dielectric contrast is achieved spatially on wavelength scales. Just as SPCs have revolutionized control over light–matter interactions by engineering the photonic density of states in space~\cite{Yablonovitch_1987, John_1987, Joannopoulos_2011, Lu_2014}, PTCs promise comparable breakthroughs from a fundamentally new perspective: a temporal one~\cite{Lustig_2018, Lyubarov_2022, Dikopoltsev_2022, Li_2023, Wang_2024, Park_2025}. However, harnessing such phenomena all-optically poses severe experimental challenges \cite{Asgari_2024, Khurgin_2020, Hayran_2022, Saha_2023, Hayran_2025}, as it requires order-unity modulation depths of a material's optical properties on ultrafast timescales comparable to the light cycle, a regime that has remained elusive to date.\\
Here, we demonstrate the first all-optical realization of a photonic time crystal, achieved with a surface plasmon cavity metamaterial operating at Terahertz (THz) frequencies. We demonstrate strong (near-unity) and coherent (sub-optical cycle) periodic driving of the plasmonic metamaterial enabled by field-induced dynamical modulation of the carriers’ kinetic energy and effective mass—reaching up to 80\% of their rest mass. Our spectroscopic measurements reveal a transition into the PTC regime mediated by an exceptional point, at which two Floquet-driven optical eigenmodes coalesce. In the PTC regime, emergent gain is shown to reduce plasmonic losses by more than 50\% \cite{Kiselev_2024, Feinberg_2025} and we predict plasmonic lasing to be within experimental reach. These results pave the way for temporal engineering of losses and light-matter interactions in plasmonic systems, and  establish a robust new platform for time-domain photonics.

\end{abstract}

\section*{Introduction}
\begin{figure*}[t]
	\centering % Centrer l'image
	\includegraphics[width=\textwidth]{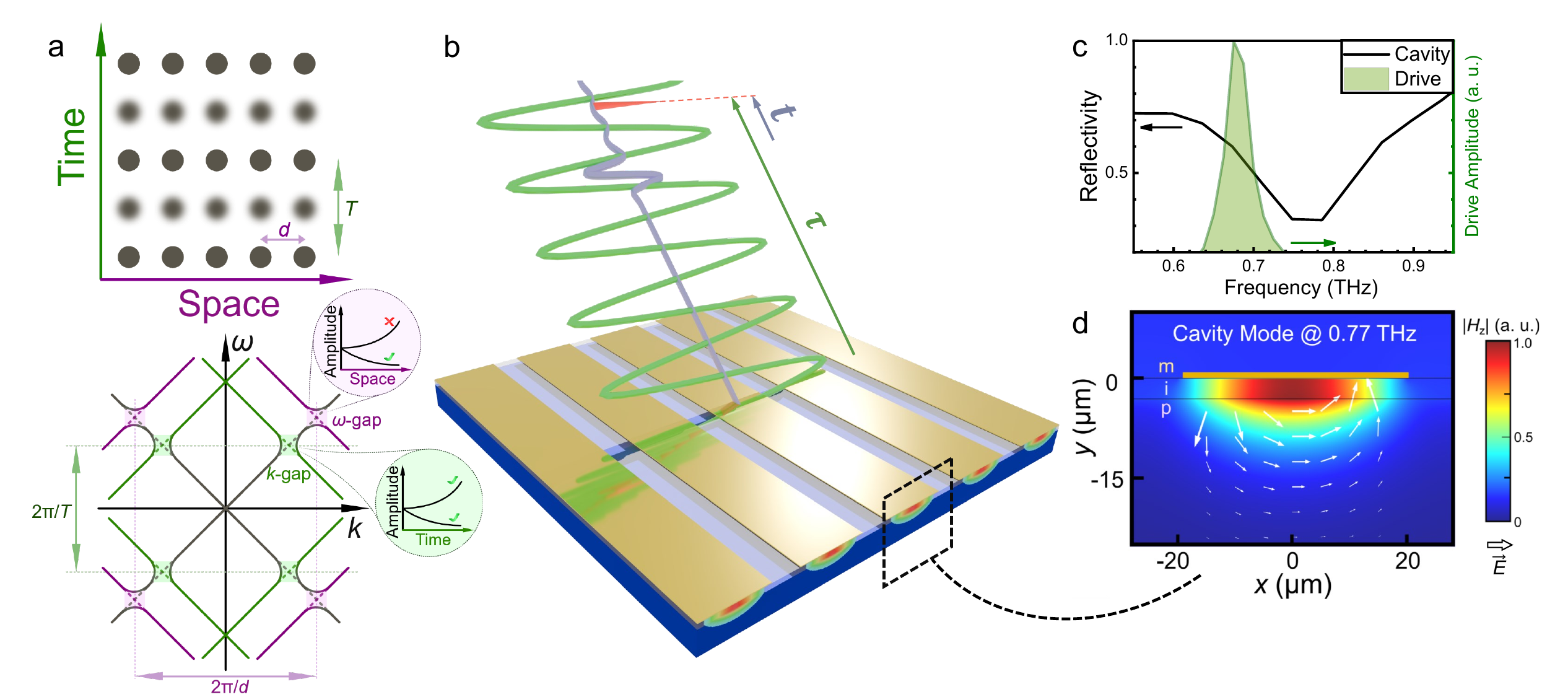}
	\caption{\textbf{a)} Conceptual representation of a combined SPC and PTC in real space (top panel) and reciprocal space (bottom panel). Periodicity in space (resp. time) generates replicas (purple and green lines) of the bare photonic band structure (dark grey lines) translated along the momentum (resp. frequency) axis. Crossings between the bare dispersion and the spatial (resp. temporal) replicas, as indicated by the dashed lines, open frequency (resp. momentum) gaps in the resulting band structure (solid lines). Insets show allowed and/or forbidden wave solutions inside the two types of gaps. \textbf{b)} Plasmonic metamaterial under consideration: a spatially periodic lattice of metal/insulator/plasmonic (m/i/p) cavities (m=Au, i=$\mathrm{Si_3N_4}$, p=InSb). The time and frequency resolved response of the metamaterial to a multi-cycle periodic drive (green pulse) is obtained by employing a broadband probe pulse (purple) that is electro-optically sampled (red pulse).  \textbf{c)} Equilibrium power reflectivity of the metamaterial at 290K (solid black curve) and normalized amplitude of the Fourier transform of the multi-cycle driving field (green)  \textbf{d)} Electric field (white vectors) and magnetic field (color bar) profiles of the cavity mode at the resonance frequency of 0.77 THz}
	\label{fig:1}
\end{figure*}

Dynamic control of optical systems in time via periodic modulation, or Floquet driving, has recently emerged as a powerful paradigm to photonic engineering. For a modulation frequency $\omega_d$ that is much smaller than the light frequency of interest $\omega$ $(\omega_d\ll \omega)$, this approach has been used to engineer a discrete synthetic dimension from the sidebands ladder at $\omega\pm n \omega_d$ ($n$ being an integer), a concept formalized within the framework of four-dimensional or Floquet metamaterials \cite{Engheta_2020, Yin_2022, Galiffi_2022, Sapienza_2023, Engheta_2023, Horsley_2023, Sisler_2024}. When the modulation and light frequencies become commensurate together $(\omega_d\approx 2\omega)$, qualitatively new phenomena emerge that enable photonic band structures to be engineered dynamically through a temporal approach, an aspect which is captured by the recently introduced concept of photonic time crystals (PTCs)  \cite{Asgari_2024, Lustig_2018, Lyubarov_2022, Dikopoltsev_2022, Li_2023, Wang_2024, Park_2025,Kiselev_2024, Feinberg_2025, Pan_2023, Hayran_2024, Khurgin_2024}. It should be noted that PTCs are periodically driven optical systems that do not break the discrete time-translational invariance imposed by the drive and should thus not be confused with many-body time crystals whose physics originate specifically from the breaking of time-translational invariance \cite{Wilczek_2012, Sacha_2017}. As depicted in Figure 1a, PTCs can be viewed as the temporal analogues of spatial photonic crystals (SPCs) \cite{Yablonovitch_1987, John_1987, Joannopoulos_2011, Lu_2014} by permuting space and time coordinates on one hand and momentum and frequency on the other. In SPCs (resp. PTCs), spatial (resp. temporal) periodicity generates band replicas along the momentum (resp. frequency) axis and whose hybridization with the original band opens frequency (resp. momentum) gaps ($\omega$-gaps, resp. $k$-gaps) near their crossings.
Wave-behavior within or in the vicinity of these gaps is key in governing light–matter interactions and differ fundamentally between the two cases. Inside the gaps, wave solutions reduce to exponentially decaying or growing modes — spatially in SPCs and temporally in PTCs (Fig. 1a). In SPCs, only the decaying solution is physically allowed, a consequence of their time-invariant and energy-conserving nature, leading to inhibition of spontaneous emission \cite{Yablonovitch_1987}. In contrast, the non-equilibrium nature of PTCs allows exponentially growing modes in time, enabling light emission and amplification \cite{Lyubarov_2022}.
This unique property makes temporally engineered band structures highly promising for novel forms of spontaneous emission control \cite{Dikopoltsev_2022, Park_2025}, overcoming electromagnetic bounds inherent to equilibrium systems \cite{Hayran_2024}, and providing non-resonant and tunable gain, lasing, and frequency conversion with reduced phase-matching requirements compared to conventional nonlinear optical methods \cite{Lyubarov_2022, Khurgin_2024}.\\
However, all-optical realizations of PTCs have proven to be a challenging task \cite{Asgari_2024, Khurgin_2020, Hayran_2022, Saha_2023, Hayran_2025}. Much like SPCs, which require a large dielectric contrast on wavelength scales to engineer photonic band structures, PTCs demand order-unity relative modulation depths of a material’s optical properties (e.g., $\delta n/n \approx 1$, where $n$ is the refractive index), together with periodic and coherent ultrafast modulations on timescales comparable to the light cycle, two conditions that remain challenging to achieve simultaneously. Consequently, experimental implementations have so far been confined to electrical circuits \cite{Reyes_Ayona_2015, Wang_2023, Lee2025}, leaving open the challenge of achieving PTCs beyond the microwave regime. \\

Here, we demonstrate the first all-optical realization of a PTC, operating at Terahertz (THz) frequencies and achieved with a surface plasmon cavity metamaterial. Using a multi-cycle THz-frequency light field of modest amplitude ($\approx 40$kV/cm), we periodically drive the plasmonic metamaterial and achieve strong (near-unity) and coherent (sub-optical-cycle) parametric modulation of its resonance, an effect which arises from dynamical variations in the carriers’ effective mass and reaching up to 80\% of their rest mass. Via spectroscopic measurements, we uncover the transition to the PTC regime as a function of time which manifests, in particular, as a substantial decrease of the plasmonic losses by more than 50\%, opening the door to the temporal engineering of plasmonic losses and plasmonic lasing in this system~\cite{Kiselev_2024,  Feinberg_2025, Kiselev2024}.

\section*{Experiment}
To demonstrate our approach, we take a conceptually simple metamaterial structure, namely a periodic lattice of sub-wavelength metal/insulator/plasma cavities, where the plasmonic material corresponds to the small band-gap semiconductor InSb  \cite{Aupiais_2023}. It is shown in Fig. \ref{fig:1}b together with the experimental configuration. 
\begin{figure*}[!ht]
	\centering % Centrer l'image
	\includegraphics[width=\textwidth]{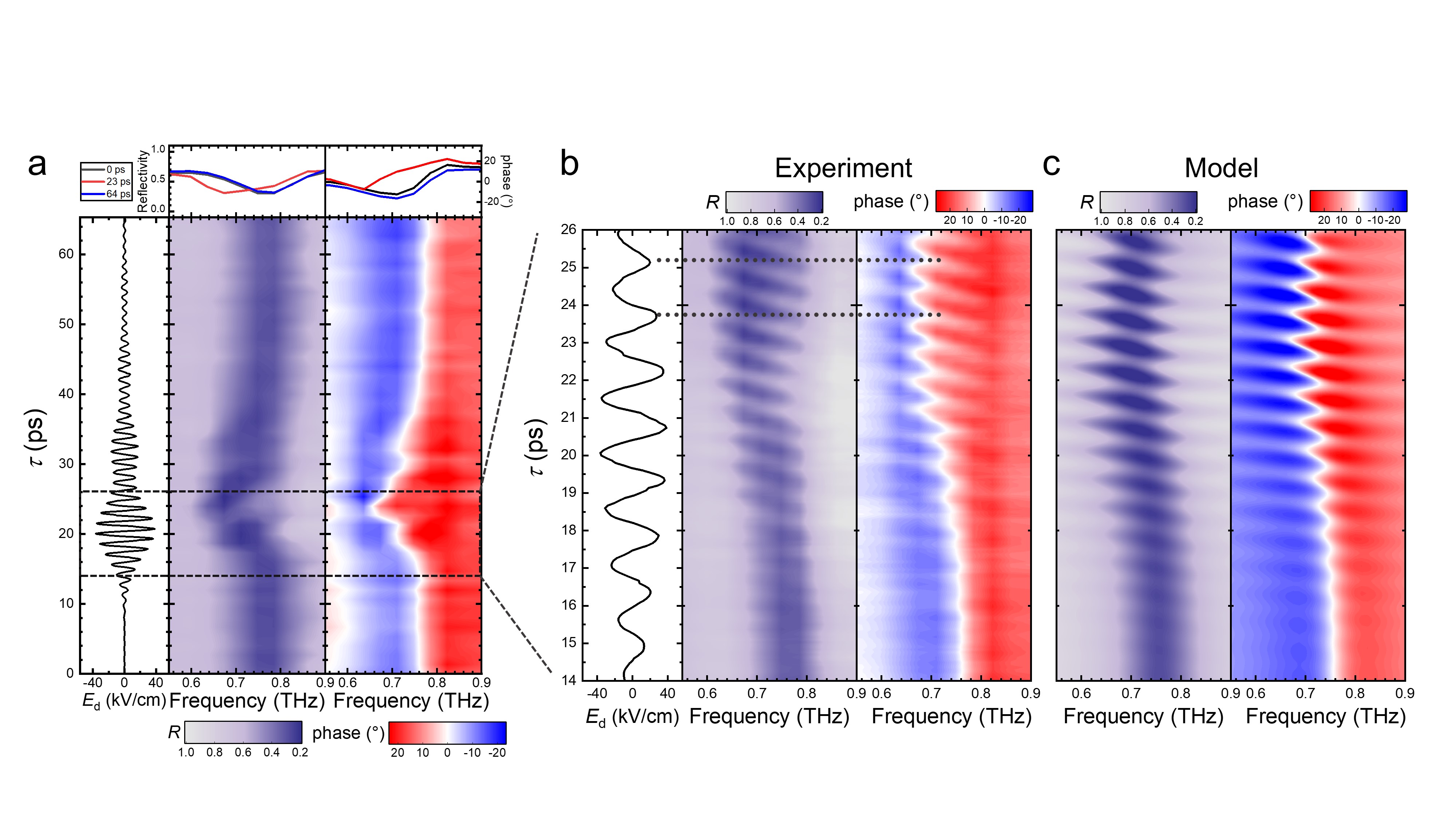}
	\caption{Temporal response of the metamaterial under a multicycle drive with a peak field of 40kV/cm: frequency-resolved power reflectivity (central panels) and phase (right panels) spectra as a function of time $\tau$ during the drive (left panels) \textbf{a)} Coarse grained dynamics measured over the complete duration of the drive \textbf{b)} Close up on the dynamics in the vicinity of the peak driving field with sub-cycle temporal resolution. \textbf{c)} Comparison of the sub-cycle dynamics shown in b) obtained from a parametrically driven cavity model (see text for details).}
	\label{fig:2}
\end{figure*}
The equilibrium power reflectivity of the metamaterial, shown in Fig. 1c, exhibits an absorption dip centered at $0.77\rm{THz}$ with a full width at half maximum of $\sim 0.18\rm{THz}$. It corresponds to the resonance of the fundamental $\lambda/2$ mode of the sub-wavelength Fabry-Perot surface plasmonic cavity, whose mode profile is displayed in Fig. 1d.
The periodically driven regime is investigated using a narrowband THz pump pulse centered at $0.69\rm{THz}$—slightly below the equilibrium resonance frequency—with a peak field amplitude of approximately 40kV/cm. The system’s dynamics during the drive is monitored by a broadband THz probe pulse, $\Tilde{E}(\tau, t)$, where $\tau$ denotes the delay between the pump and the electro-optic sampling beam, and $t$ represents the internal delay between the sampling beam and the THz probe pulse, as defined in Fig. 1b. For each time delay $\tau$, the frequency-resolved response of the metamaterial is obtained by performing a Fourier transform over the internal delay $t$.\\
Fig. 2a-b displays the temporal evolution of the power reflectivity and phase of the metamaterial's response, as measured experimentally. Fig. 2a shows the metamaterial's response across the full duration of the drive with coarse temporal resolution, while Fig.2b provides a close-up view in the vicinity of the peak of the driving field with sub-cycle temporal resolution.
During the interaction, we observe a pronounced redshift of the cavity resonance, with an amplitude that closely tracks the envelope of the driving field. At large positive delays-when the driving field has almost vanished -the metamaterial response returns to its equilibrium state without any signature of a long-lived dynamics nor heating, suggesting a coherent interaction between the drive and the metamaterial.
The measurement at sub-cycle temporal resolution in Fig. 2b fully confirms this picture, revealing pronounced and periodic modulations in both reflectivity and phase, which are driven on a timescale comparable to the optical cycle. Following an initial buildup of the response, the metamaterial reaches a quasi–steady-state characterized by 1) a red-shift of the metamaterial resonance frequency on average and 2) forced oscillations of the resonance parameters around this state. As can be seen from the dotted lines, these oscillations occur twice per optical cycle, corresponding to the second harmonic of the drive frequency. Below we model this system from basic principles to gain physical insights into its dynamics and quantify the effect of the time-modulation on the metamaterial optical properties.

\section*{Model}

We attribute the temporal modulations of the surface plasmon cavity resonance to dynamical variations in its kinetic inductance through the carriers’ effective mass $m^*(\tau)$, which we write $ m^*(\tau) = m_0^* + \delta m^*(\tau)$, where $m_0^*$ denotes the carriers' rest effective mass. Such variations are known to arise from non-parabolic conduction bands in narrow band-gap semiconductors \cite{Houver_2019, Soranzio_2024} and have also been discussed in the context epsilon-near-zero plasmonic materials operating at optical frequencies \cite{Khurgin_2020}. \\
Under periodic driving at frequency \( \omega_\mathrm{d} \), the electrons' momentum $p(\tau)$ oscillates as
\(
p(\tau) = p_0 \cos(\omega_\mathrm{d} \tau)
\). For a non-parabolic and isotropic conduction band given by the Kane model \cite{Kane_1957}, this results in effective mass modulations at twice the drive frequency:
\begin{equation}
\delta m^*(\tau)=
 \alpha p^2(\tau) =  \frac{\alpha}{2}{p_0}^2+\frac{\alpha}{2}{p_0}^2\cos(2\omega_\mathrm{d} \tau)
 \label{mass}
\end{equation}
where \( \alpha \) accounts for the deviation from a purely parabolic band dispersion ($\alpha>0$ here, see methods).
The resulting oscillation of the plasmonic cavity resonance can be qualitatively understood as follows. Using an LC-resonator model for the surface plasmonic cavity, the resonance frequency is given by
\(
\omega_\mathrm{c} = \frac{1}{\sqrt{LC}},
\)
where \( L \) and \( C \) are the inductance and capacitance of the mode, respectively. For surface plasmons, the inductance \( L=L_G+L_K \) includes contributions from both a geometric inductance \( L_G \), arising from retardation effects, and a kinetic inductance \( L_K \propto \frac{m^*}{ne^2} \), which represents part of the mode energy that is stored into the kinetic energy of the carriers (with \( n \) the carrier density and \( e \) the charge) \cite{Staffaroni_2012}.
The kinetic inductance can be separated into an equilibrium and a time-varying part: \(
L_K = L_K^0 + \delta L_K(\tau)
\). Rigorously, it can be shown that \(
\delta L_K(\tau)/L_K^0 = \beta \delta m^*(\tau)/m^*_0 \) where the time-independent parameter $0\leq\beta\leq1$ reaches unity only in the case of negligible geometric inductance and is lower otherwise (see methods and supplementary material). 
The resulting modulation of the resonance frequency is:
\begin{equation}
\label{cavity}
\omega_\mathrm{c}(\tau)\approx \omega_\mathrm{c}^0 \left(1 - \frac{1}{2} \frac{L_K^0}{L_G + L_K^0} \beta \frac{\delta m^*(\tau)}{m^*_0} \right)
\end{equation}
where
\(
\omega_\mathrm{c}^0 = \frac{1}{\sqrt{(L_G + L_K^0) C}}
\)
is the equilibrium resonance frequency.
This equation has a simple physical interpretation: since modulation of the cavity frequency originates from variations of the kinetic inductance via the carriers' effective mass, its depth is controlled by the relative contributions of the kinetic and geometric inductance to the cavity mode. The presence of a geometric inductance $L_G$, which is unaffected by the drive, acts only to reduce the overall depth of the cavity modulation.
Taken together, equations (\ref{mass}) and (\ref{cavity}) predict the red-shift of the cavity resonance and its modulation around the shifted resonance at the second harmonic of the drive frequency, as observed experimentally.\\
Such qualitative description can be put on solid grounds starting from the Hamiltonian and canonical quantization procedures. Specifically, excitations of the surface plasmonic cavity field described by their creation and annihilation operators $\hat{a}$ and $\hat{a}^{\dagger}$ evolve according to the following Hamiltonian:
\begin{align}
\label{eq:QuantizedHamiltonianCreationOperatorMain}
\hat{H}(\tau) = \hbar \omega_{\rm c}^0 \hat{a}^{\dagger} \hat{a} - \frac{\hbar\eta(\tau)}{2} \left(\hat{a}^{\dagger} -  \hat{a}\right)^2  
\end{align}
where the time-dependent modulation parameter is given by  $\eta(\tau)=-\frac{\omega_{\rm c}^0}{2} \frac{L_K^0}{L_G + L_K^0} \beta \frac{\delta m^*(\tau)}{m^*_0}$ (see methods and supplementary material).
To capture the metamaterial response function that is measured experimentally -the reflectivity-, we implement an input-output formalism with two baths representing radiative and non-radiative (ohmic) decay channels. These account for coupling to an external electromagnetic probe and material losses, respectively. Adopting the experimental temporal conventions and using a single adjustable parameter, we present in Fig. 2c the computed time-resolved reflection coefficient $\tilde{r}(\tau, \omega)$ that best matches the experimental observations. This approach allows direct comparison between experimental and theoretical response functions, enabling inference of the time-dependent modulation of microscopic system parameters.
\begin{figure*}[!ht]
	\centering % Centrer l'image
	\includegraphics[width=\textwidth]{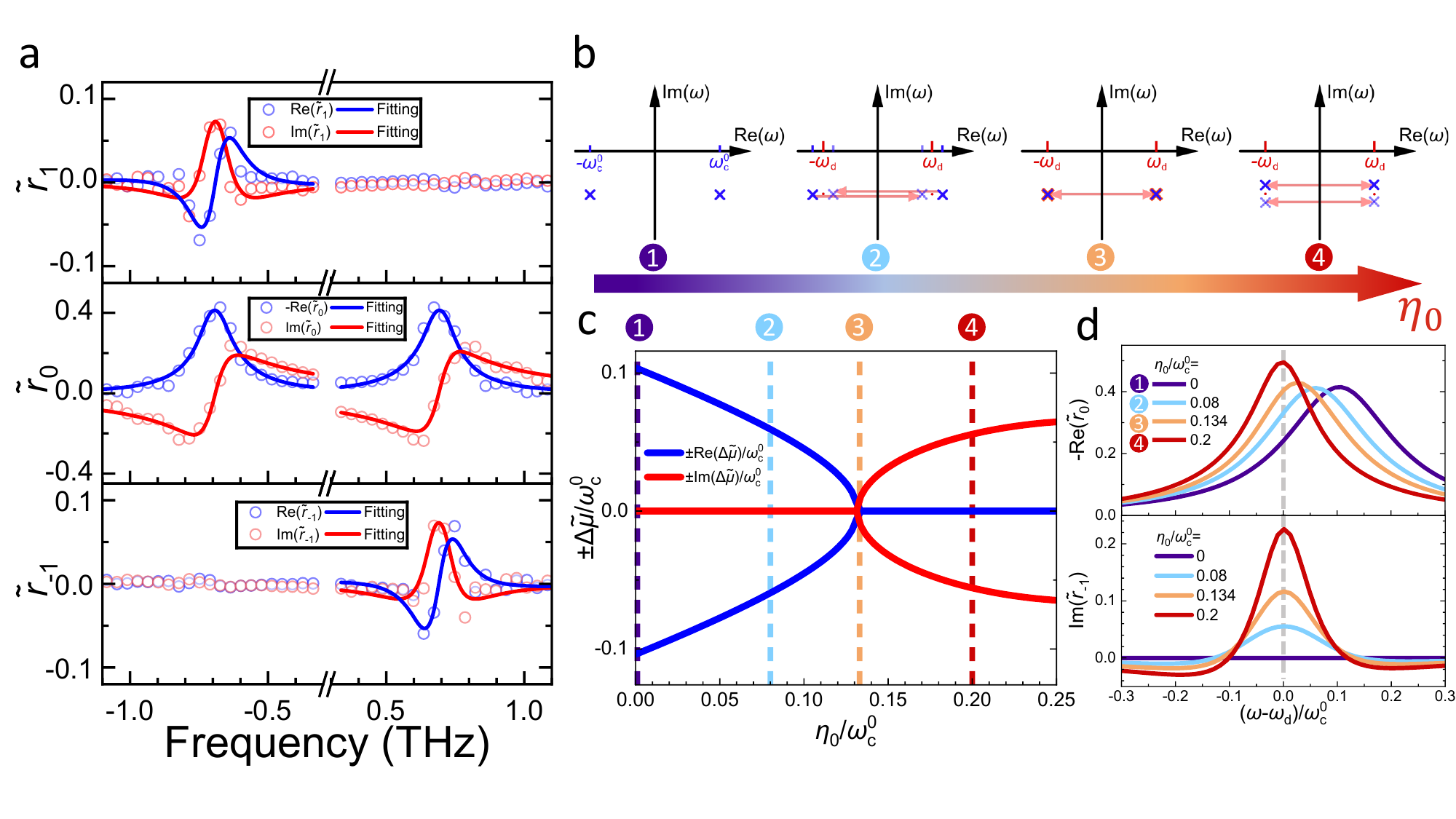}
	\caption{\textbf{a)} Real and imaginary parts of the zero-order (center panel) and first-orders (outer panels) Floquet spectra obtained from Fourier Transform of the time domain data of Fig 2b. over one drive period at $\tau\approx 24ps$. Experimental data points are shown together with a fitting of the lineshapes (see text). \textbf{b)} Evolution of the complex-frequency poles of the reflectivity, controlled by the Floquet eigenmodes, as a function of drive strength across the PTC transition. \textbf{c)} Real and imaginary parts of $\Delta\tilde{\mu}/\omega_c^0$ as a function of drive strength $\eta_0/\omega_c$ computed for parameters taken from the experiment: $\omega_d/\omega_c^0 = 0.92$, $\Gamma/\omega_c^0=0.12$. Dashed lines indicate the values of $\eta_0/\omega_c$ taken for the computation of the spectra shown in Fig. 3d. \textbf{d)} Spectra of the real part of $\tilde{r}_0(\omega)$ and imaginary part of $\tilde{r}_{-1}(\omega)$  across the PTC transition and computed for values of $\eta_0/\omega_c$ indicated by the dashed lines in panel \textbf{c}.}
	\label{fig:3}
\end{figure*}

\section*{Discussion}
We obtain a maximum modulation parameter $\eta_0/\omega_\mathrm{c}^0=\max \limits_{\tau}|\eta(\tau)|/\omega_\mathrm{c}^0=0.18 \pm 0.02$ for which the model reproduces the experimental data with very good accuracy. This corresponds to an impressive $\delta m^*(\tau)/m^*_0\approx 80\%$ dynamical effective mass modulations. This value is not limited by the maximum driving field that can be applied, which could be further increased, but rather by the onset of incoherent nonlinearities, namely impact ionization (see Supplementary Material), which starts to emerge at fields exceeding $\sim 40 kV/cm$ due to the small band-gap of the InSb plasmonic material \cite{Houver_2019, Biasco_2022}. Hence, it appears to be a physical bound for coherent effective mass modulations in this system. 
\subsubsection*{Spectroscopy of a PTC}
We next investigate the periodically driven quasi--steady-state regime 
observed at $\tau_0 \simeq 24\,\mathrm{ps}$ by analyzing the dynamics over one period of the drive $\tau \in [\tau_0, \tau_0 + 2\pi/\omega_d]$ within the framework of Floquet theory. For a system that is periodic in time with frequency $2\omega_d$, the complex reflectivity $\tilde{r}(\tau,\omega)$ can be expanded in Fourier harmonics 
at frequencies $\pm 2n\omega_d$ (see Supplementary material). In the present case, this expansion is dominated by the zeroth-order (time-averaged) component and the first-order harmonic contributions, such that:
\begin{equation}
\tilde{r}(\tau,\omega)
=
\tilde{r}_b(\omega)
+
\tilde{r}_0(\omega)
+
\sum_{n=\pm1}
\tilde{r}_n(\omega)\,e^{2in\omega_d\tau}.
\end{equation}
The time-averaged response consists of a non-resonant background $\tilde{r}_b(\omega)$ and a resonant cavity contribution $\tilde{r}_0(\omega)$, while $\tilde{r}_{\pm 1}(\omega)$ are the first-order Floquet sidebands. The latter correspond to photons whose frequency has been either up-converted, $\tilde{r}_1(\omega)$, or down-converted, $\tilde{r}_{-1}(\omega)$, by $2\omega_d$ with respect to their incident frequency $\omega$. Spectra for $\tilde{r}_0(\omega)$ and $\tilde{r}_{\pm1}(\omega)$ are presented in Fig.~3a.
Although the reflectivity satisfies $\tilde{r}(\tau,\omega)=\tilde{r}(\tau,-\omega)^{*}$ at any time $\tau$, spectra of the Floquet sidebands $\tilde{r}_{\pm1}(\omega)$ display two striking features: as opposed to $\tilde{r}_0(\omega)$, they are predominantly one-sided in frequency and their lineshapes deviate qualitatively from that expected of a simple resonance. Instead, they manifest characteristics of a second order resonance (or pole) emanating from the presence of two closely spaced resonances within the spectral linewidth (see below). As shown in Fig. 3b, these aspects can be interpreted as follows. In the absence of modulation, the cavity resonates for both positive ($\omega\approx\omega_c^0$) and negative ($\omega\approx-\omega_c^0$) frequencies within the spectral width $\Gamma$, which is described mathematically as two poles in the reflectivity located in the complex-frequency plane at $\tilde{\omega}_c^0=\omega_c^0-i\Gamma$ and $-(\tilde{\omega}_c^0)^*=-\omega_c^0-i\Gamma$ (Fig. 3b-panel (1)). The periodic drive generates Floquet replicas of these poles shifted by integer multiples of $\pm 2\omega_d$. Notably, when the drive frequency is tuned close to the cavity resonance ($\omega_d\approx\omega_c^0$), the first-order replica of the negative (resp. positive) frequency component becomes nearly degenerate with its original positive (resp. negative) frequency counterpart i.e., $-\omega_c^0+ 2\omega_d\approx \omega_c^0$ (resp. $\omega_c^0-2\omega_d\approx -\omega_c^0 $) (Fig. 3b-panel (2)). These two Floquet eigenmodes are described as two closely spaced poles, which can be written on the positive frequency side as: 
\begin{equation}
\tilde{\mu}_{\pm}
=
\tilde{\omega}_d \pm \Delta\tilde{\mu},
\end{equation}
where $\tilde{\omega}_d=\omega_d-i\Gamma $. One can show that the Floquet matrix displays PT symmetry (see Supplementary material) which, if unbroken, constrains $\Delta\tilde{\mu}$ to be real-valued ~\cite{bender1998real,ganainy2018nonhermitian}, as it is the case for weak driving, see Fig.~3c. In this regime, photons undergo repeated scattering between the positive and negative frequency components of the resonance and find a resonant pathway to get either up or down converted. Importantly, non-resonant pathways in which, for instance, positive frequency photons attempt to get up-converted are strongly inhibited, explaining why the Floquet sidebands appear predominantly one-sided (e.g. $|\tilde{r}_{1}(+\omega)|\ll|\tilde{r}_{1}(-\omega)|$). Ultimately, the Floquet sidebands probe the joint spectral density of states, i.e. the spectral overlap $\propto
\frac{1}{(\omega-\tilde{\mu}_{+})(\omega-\tilde{\mu}_{-})}$, of the two Floquet eigenmodes, which naturally accounts for their second-order pole character (see Supplementary material). \\
As the driving strength is increased, the coupling between those two eigenmodes gets stronger, their separation
$\Delta\tilde{\mu}$ decreases and eventually vanishes, at which point they coalesce at $\tilde{\omega}_d$ (Fig. 3b-panel (3)). This coalescence defines an exceptional point characterized by a spontaneous breaking of PT symmetry~\cite{koutserimpas2018parametric,bossart2021nonhermitian} and marks the transition into the PTC regime. Beyond this point, the difference between the two Floquet eigenmodes $\Delta\tilde{\mu}$ become purely imaginary
%the resonances split in their decay rates, 
\begin{equation}
\tilde{\mu}_{\pm}=\tilde{\omega}_d \pm i\Delta\Gamma
\end{equation} 
leading to reduced damping in one of the two modes, which is a hallmark of the PTC phase (Fig. 3b-panel (4)) \cite{Asgari_2024, Feinberg_2025}.\\
In Fig.~3c-d, using Floquet analysis with periodic drive $\eta(\tau)=-\eta_0 \cos^2(\omega_\mathrm{d} \tau)$, we compute the evolution of $\Delta\tilde{\mu}/\omega_c^0$ (Fig. 3c) and of the spectral lineshapes expected across this transition  (Fig 3d), from which two important conclusions can be drawn: 1) the spectra evolve smoothly across the PTC transition and 2) the lineshapes' characteristics of $\tilde{r}_{0}(\omega)$ and $\tilde{r}_{\pm1}(\omega)$ (both real and imaginary parts) encode detailed information about the poles' structure and the system's parameters, notably the spectral locations of $\tilde{\mu}_{\pm}$ which can be obtained through fitting of the experimental lineshapes (see Methods and Supplementary material).

\begin{figure*}[!ht]
	\centering % Centrer l'image
	\includegraphics[width=\textwidth]{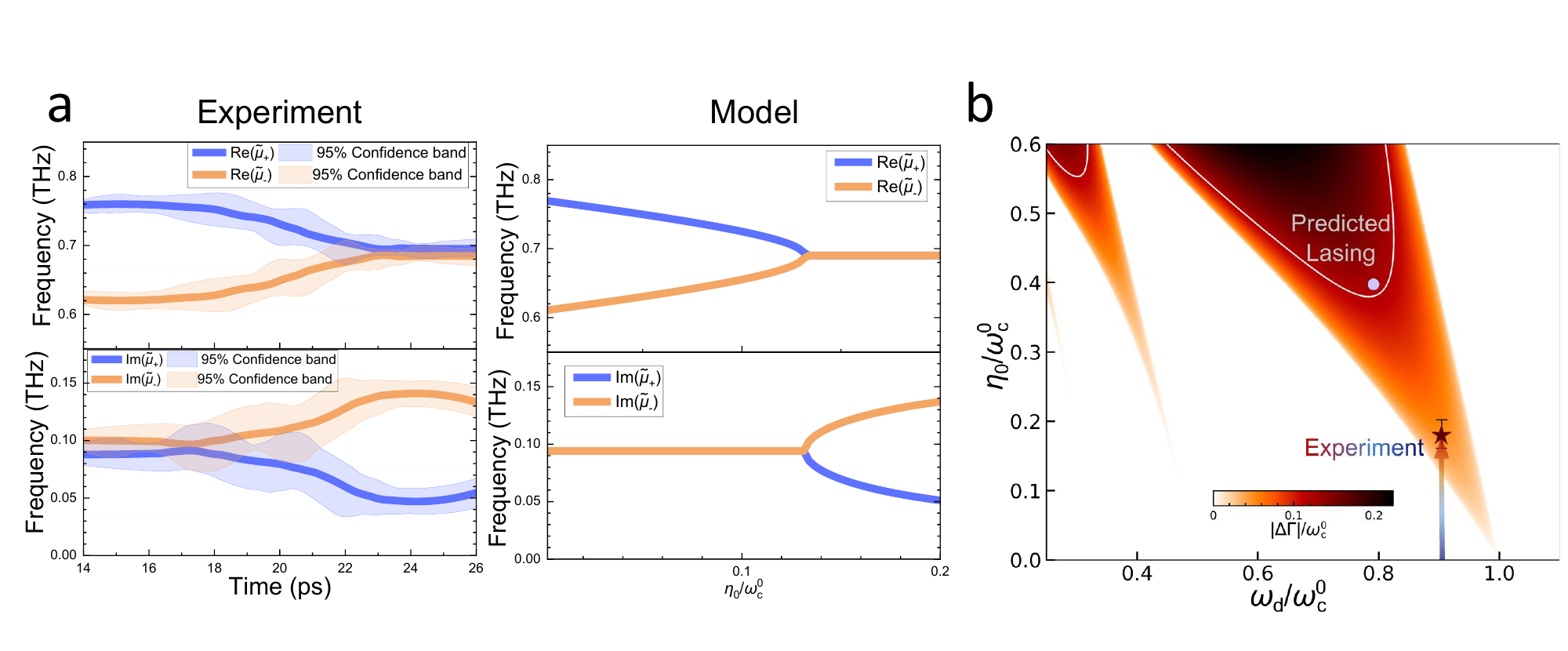}
	\caption{\textbf{a} Left panel: Temporal evolution of the Floquet eigenvalues $\tilde{\mu}_{\pm}$ (solid lines) shown with 95\% confidence interval Right panel: Evolution of the Floquet eigenvalues $\tilde{\mu}_{\pm}$ as a function of drive strength $\eta_0/\omega_c^0$ computed from the model for parameters taken from the experiment.  \textbf{c)} Stability diagram of the driven system: parametric gain $\Delta \Gamma$ as a function of drive frequency $\omega_\mathrm{d}$ and modulation strength $\eta_0$ relative to the equilibrium cavity frequency $\omega_\mathrm{c}^0$. The colored area where $\Delta \Gamma \neq 0$ signals the PTC regime. The solid contour line defined by $\Gamma/\omega_\mathrm{c}^0=0.12$ marks the threshold for plasmonic lasing given the  measured equilibrium losses $\Gamma$.}
	\label{fig:4}
\end{figure*}

\subsubsection*{Temporal transition to the PTC regime}
Below, we unambiguously confirm the transition to the PTC regime by extracting the spectral locations of the Floquet eigenmodes as a function of time $\tau$. Treating the drive amplitude as a slowly varying function of time with respect to the fast modulations at $\pm 2\omega_d$, this is achieved by performing a short term Fourier Transform of the time-domain data (Fig. 2b) to extract the slowly varying components $\tilde{r}_0(\tau,\omega)$ and $\tilde{r}_1(\tau,\omega)$ and performing a fit of the lineshapes to extract the spectral location of the poles. The results are presented in Fig 4a. As a function of time, we observe the pole coalescence and the transition to the PTC regime at around $\tau\approx 21ps$ before the poles acquire different imaginary part. The experimental results are seen to be in excellent agreement with predictions of our model, notably in the magnitude of the effect obtained. Within the PTC regime, the emergence of gain is directly reflected in the reduction of the imaginary part of one of the poles. Experimentally, this corresponds to a linewidth narrowing of approximately $\Delta\Gamma\approx40GHz$ from an initial equilibrium linewidth 
$\Gamma=\gamma_r+\gamma_{nr}$, with $\gamma_{nr}\approx70GHz$ and $\gamma_{r}\approx20GHz$ being the non radiative and radiative losses respectively. It indicates that the non-radiative plasmonic losses $\gamma_{nr}$ are reduced by more than 
50\%. To the best of our knowledge, this constitutes the first demonstration of plasmonic loss reduction induced by a parametric drive, opening a new route for plasmonic loss engineering based on a temporal approach \cite{Kiselev_2024, Feinberg_2025}.
\subsubsection*{Towards plasmonic lasing}
Finally, we place these results in the broader context of this platform's parameters and discuss the possibility of plasmonic lasing in this system, which may occur when the gain produced by the drive overcomes the total losses ($\Delta\Gamma > \Gamma $). To that end, we compute in Fig. 4b its phase diagram as a function of drive frequency and strength. The colored area marks the PTC phase with the emergence of gain ($\Delta\Gamma\neq 0$), while the white area corresponds to the absence of gain ($\Delta\Gamma= 0$). The current experimental realization of Fig 4a are reported as colored dots on this diagram. We predict that plasmonic lasing with this platform is within experimental reach with an improved cavity design and fine tuning of the drive parameters. A strategy for this consists in increasing the ratio $L_K^0/L_G$ by decreasing the size of the plasmonic cavities so as to operate in the 'electrostatic' regime where retardation effects vanish and kinetic ones dominate \cite{Staffaroni_2012,Greffet_2012}. For such an electrostatic cavity ($\beta=1, L_K^0\gg L_G$, $\eta_0/\omega_\mathrm{c}^0=0.4$, white dot in Fig 4b), this enhances parametric effects as compared to the present cavity design ($\beta\approx 0.65, L_K^0/L_G\approx 2.4$, $\eta_0/\omega_\mathrm{c}^0\approx 0.18$, star in Fig 4b), and may ultimately allow to cross the threshold for lasing, as indicated by the white solid line shown in Fig.~4.
 \\

\section*{Conclusion}
In conclusion, we report the first all-optical realization of a photonic time crystal, operating at THz frequencies and implemented with a plasmonic metamaterial. The underlying mechanism relies on large and ultrafast dynamical modulations of the carriers' effective mass sustaining the plasmonic resonance. The transition to the PTC regime, evidenced by direct spectroscopic measurements and fully supported by a theoretical analysis based on Floquet theory, manifests through the coalescence of two Floquet-driven optical modes followed by the onset of plasmonic gain. This work paves the way towards the temporal engineering of dissipation and light-matter interactions in plasmonic systems and establishes a solid framework for the realization and spectroscopic identification of PTC phenomena in future implementations.

\section*{Acknowledgments}

Y.L. acknowledges support from Agence Nationale de la Recherche (grant n$^{\circ}$ ANR-23-CE30-0030) and Region Ile-de-France in the framework of DIM QuanTiP. M.S. acknowledges financial support
from the ERC consolidator Grant No. 101002955 - CONQUER. T.G. acknowledges support from Chinese Scholarship Council (CSC202206210114). G.M.A. acknowledges funding from the European Union’s Horizon 2020 research and innovation programme under the Marie Skłodowska-Curie Grant Agreement No. 101146870 – COMPASS. Parts of this research were carried out at ELBE at the Helmholtz-Zentrum Dresden-Rossendorf e.V., a member of the Helmholtz Association.

\section*{Methods}

\subsection*{Sample manufacturing}

A $\langle 100 \rangle$-oriented, nominally undoped InSb crystal with dimensions of $5~\mathrm{mm} \times 5~\mathrm{mm} \times 0.5~\mathrm{mm}$ is used as the starting material. A $\sim3~\mu\mathrm{m}$-thick $\mathrm{Si}_3\mathrm{N}_4$ insulating layer is first deposited via plasma-enhanced chemical vapor deposition. The top metallic pattern is then defined through photolithography, followed by Ti/Au (20 nm/200 nm) evaporation and lift-off. The resulting metamaterial structure consists of periodic metallic stripes with a width of $s = 41~\mu\mathrm{m}$ and spacing $a = 16~\mu\mathrm{m}$, yielding a total period of $d = s + a = 57~\mu\mathrm{m}$ and follows the design reported in a previous work~\cite{Aupiais_2023}.

\subsection*{THz pump-probe spectroscopy}
The driving beam is generated by the superradiant Terahertz source TELBE at the ELBE accelerator (Helmholtz-Zentrum Dresden-Rossendorf). It produces tunable narrow-band and multi-cycle pulses from $0.1$ to $3$ THz and a selectable repetition rate from 10 to 250 kHz. For this experiment, the source was tuned at a central frequency of $0.69~\mathrm{THz}$ with a $50~\mathrm{kHz}$ repetition rate. The broadband THz probe is generated by a photo-conductive emitter (LaTera 10/10, HZDR Innovation) driven by a $100~\mathrm{kHz}$ ultrafast Ti:Sapphire amplifier system synchronized to TELBE via a stabilized fiber link. THz pulse measurements are achieved via a balanced detection and electro-optical sampling in a $100~\mu\mathrm{m}$-thick $\langle 110\rangle$ ZnTe crystal. THz pump and probe, combined with a HRFZ Silicon plate, are co-propagating and co-polarized and are focused onto the sample at $45^\circ$ incidence angle with s-polarization in a spot of FWHM diameter around $900~\mu\mathrm{m}$. All measurements are performed at room temperature in a purged atmosphere. \\
Time-domain windowing is applied to isolate the main THz pulse from other time-delayed reflections and is Fourier transformed to obtain power reflectivity and phase spectra. The reflection coefficient is obtained based on a self-referencing method by normalizing the TM response to the TE response of the sample: TM-polarized THz fields (perpendicular to the stripes) excite cavity modes, while TE-polarized fields (parallel to the stripes) are fully reflected and are used as a reference. To avoid nonphysical frequency-independent contributions to the signal, as can arise from variations in the generation or balancing conditions of the detection, power reflectivity spectra are scaled to ensure reaching the same limit in the low frequency part of the spectra ($0.2$-$0.45~\mathrm{THz}$). 

\subsection*{Theoretical model}
The collective motion of electrons leading to the surface plasmon cavity resonance is described as a $LC$ oscillator.
The Hamiltonian can be expressed as:
\begin{align}
H=\frac{1}{2C}Q^2+\frac{1}{2}LI^2
\label{eq:H} 
\end{align}
Here, $Q$ and $I$ are collective variables describing the dynamics of the surface plasmon cavity field. They are associated respectively with the charge displacement and the plasmonic current and are related by charge conservation ($\dot{Q}=I$). $C$ and $L$ parametrize the capacitive and inductive parts of the energy of the plasmon field.\\ \\
The inductance $L$ has two contributions \cite{Staffaroni_2012}:
\begin{itemize}
    \item a geometric inductance $L_G$ arising from retardation effects of the electromagnetic field
    \item a kinetic inductance $L_K\propto\frac{m^*}{ne^2}$ characterizing energy that is stored into the kinetic energy of the carriers ($m^*$,$n$ and $e$ are the mass, density and charge of the carriers).\\ We decompose it as a sum of the rest kinetic inductance of the carriers $L_K^0$ and a time-dependent contribution $\delta L_K(\tau)$ arising from effective mass modulations: $L_K=L_K^0+\delta L_K(\tau)$. 
\end{itemize}
The total inductance is the sum of these different contributions:
\begin{align}
L=L_G+L_K^0+\delta L_K(t)
\label{eq:L}  
\end{align}
With these definitions, the Hamiltonian writes :
\begin{align}
H(\tau)=\frac{1}{2C}Q^2+\frac{1}{2}(L_G+L_K^0)I^2+\frac{1}{2}\delta L_K(\tau)I^2
\label{eq:H} 
\end{align}  \\ \\
After quantization, this can be expressed with ladder operators $ \hat{a}$ and $\hat{a}^{\dagger}$ describing excitations of the surface plasmon cavity field as: 
\begin{align} \label{eq:QuantizedHamiltonianCreationOperator}
\hat{H}(\tau) = \hbar \omega_{\rm c}^0 \hat{a}^{\dagger} \hat{a} - \frac{\hbar\eta(\tau)}{2} \left(\hat{a}^{\dagger} -  \hat{a}\right)^2
\end{align}
where the modulation parameter $\eta(\tau)$ is:
\begin{align}
\eta(\tau)=-\frac{\omega_{\rm c}^0}{2}\frac{\delta L_K(\tau)}{L_G+L_K^0}
\label{eq:eta1} 
\end{align} 
The modulation parameter can be also expressed as a function of the effective mass modulations as:
\begin{align}
\eta(\tau)=-\frac{\omega_{\rm c}^0}{2}\frac{L_K^0}{L_G+L_K^0}\beta\frac{\delta m(\tau)}{m^*_0}
\label{eq:eta2} 
\end{align} 
where $0\leq\beta\leq1$ is a time-independent factor that links the variations of the kinetic inductance to the variations of the effective mass: $\frac{\delta L_K(\tau)}{L_K^0}=\beta\frac{\delta m^*(\tau)}{m^*_0}$. Its computation is outlined below.
Equation (\ref{eq:QuantizedHamiltonianCreationOperator}) is supplemented with an input-ouput formalism to compute the reflection coefficient, as detailed in the supplementary material.

\subsection*{Model parameters}
Computation of the parameters entering equations (\ref{eq:eta1}) and (\ref{eq:eta2}) is outlined here and explained in more detail in the supplementary material. These calculations are based on the determination of the cavity resonance frequency $\omega_{\rm c}$, obtained by solving the dispersion relation of a truncated metal/insulator/plasmonic waveguide \cite{Aupiais_2023}, when parameters are varied. Notably, the resonance frequency depends, among other geometrical and optical parameters of the constitutive elements of the metamaterial, on the plasma frequency of the plasmonic medium supporting the resonance: $\omega_{\rm p}=\sqrt{\frac{ne^2}{\epsilon_{0}\epsilon_{\infty}m^*}}$, where $n$ and $e$ are the density and charge of the carriers, $m^*$ their effective mass, $\epsilon_{0}$ the vacuum permittivity and $\epsilon_{\infty}$ the high-frequency permittivity of the plasmonic medium. Therefore, the ratio of kinetic to geometric inductance $ L_K^0 / L_G $ at equilibrium can be determined from the resonance frequencies $\omega_\mathrm{c}^0 $ and $\omega_0 $ obtained by varying the plasma frequency for two otherwise identical cavities: 1) with equilibrium plasma frequency, yielding $
\omega_{\rm c} = \frac{1}{\sqrt{(L_K^0 + L_G)C}}$ and 2) with infinite plasma frequency, for which the kinetic inductance vanishes while the geometric inductance and capacitance remain unchanged, yielding $\omega_0 = \frac{1}{\sqrt{L_G C}}$.
Similarly, the dependence of kinetic inductance with the effective mass, allowing to determine $\delta L_K$ and $\beta$, is obtained through a variation of the effective mass that enters the plasma frequency, all other parameters being fixed.
We point out that these computations are parameter-free and obtained from experimental measurements of the geometrical and optical properties of the constitutive elements of the metamaterial, as detailed in \cite{Aupiais_2023}. 
\subsection*{Temporal dynamics (Figure 2)}
To compute the temporal response of the system to the experimentally measured incident driving field $E_d(\tau)$ shown in Fig.~\ref{fig:2}, we determine $\eta(\tau)$ from equation (\ref{eq:eta2}) as follows. We consider a non-parabolic band structure given by the Kane model suitable for the description of small-band gap semiconductors \cite{Kane_1957} where the electronic dispersion relation $\varepsilon(p)$ satisfies $\varepsilon(p) (1+\alpha \varepsilon(p) ) =\frac{p^2}{2m^*_0} $ and $\alpha$ is the parameter that accounts for non-parabolicity. The effective mass $m^*(p)$ is defined by $m^*(p)=p\left( \frac{d \varepsilon}{dp}\right)^{-1}$, which allows to express it as: $m^*(p)=m^*_0(1+2\alpha \varepsilon(p))$ \cite{Zawadzki_1974}.
Keeping only the leading order term $\propto p^2$ in the electronic dispersion relation, we parametrize the effective mass as $m^*(p)=m^*_0(1+2\alpha \frac{p^2}{2m^*_0})$, so that  $\frac{\delta m^*(p)}{m^*_0}=\frac{\alpha p^2}{m^*_0}$.
We solve the dynamics of the electrons' momentum $p(\tau)$ in the resonator according to: 
\begin{align}
\frac{d^2 p}{d\tau^2} + 2\Gamma \frac{d p}{d\tau}+ {\omega_\mathrm{c}^0}^2 p &=   -e \frac{d}{d\tau}  (fE_\mathrm{d}(\tau))
\label{eq:SemiclassicalBloch}  
\end{align}
where the resonator frequency is kept fixed for simplicity and where $f$ accounts for the fraction of the incident field $fE_\mathrm{d}(\tau)$ that effectively enters the plasmonic medium and couples to the electrons' momentum. It is the only free parameter of our model.
Solving this equation for $p(\tau)$, we determine $\eta(\tau)=-\frac{\omega_{\rm c}^0}{2}\frac{L_K^0}{L_G+L_K^0}\beta\frac{\alpha p(\tau)^2}{m^*_0}$ and compute the quantum dynamics eq. (\ref{eq:QuantizedHamiltonianCreationOperator}) with an input-output formalism to obtain the reflection coefficient plotted in Fig. 2c.\\
\noindent The parameters used in the computation are listed in Table 1.
\begin{table}[htbp]
\centering
\label{params}
\caption{Model Parameters}
\resizebox{\columnwidth}{!}{
\begin{tabular}{@{}llll@{}}
\hline
\textbf{Parameter Name} & \textbf{Notation} & \textbf{Value (SI Unit)} & \textbf{Source} \\
\hline
Non-parabolicity parameter & $\alpha$ & $5.6~\mathrm{eV}^{-1}$ & \cite{Zawadzki_1974} \\
Radiative power decay rate & $\gamma$ & $2\pi \times 0.042~\mathrm{THz}$ & Experimental fit \\
Non-radiative power decay rate & $\gamma'$ & $2\pi \times 0.141~\mathrm{THz}$ & Experimental fit \\
Total field decay rate & $\Gamma=(\gamma+\gamma')/2$ & $2\pi \times 0.091~\mathrm{THz}$ & Experimental fit \\
Cavity resonance frequency & $\omega_\mathrm{c}^0$ & $2\pi \times 0.77~\mathrm{THz}$ & Experimental fit\\
Effective drive field & $f$ & $27\%$ & Free parameter \\
Inductance ratio & $L_K^0/L_G$ & $2.4$ & Computed \\
Inductance to effective mass variations & $\beta$ & $0.65$ & Computed\\
\hline
\end{tabular}}
\end{table}
\subsection*{Spectroscopic analysis based on Floquet theory (Figures 3 \& 4)}
\subsubsection*{Floquet theory}
Under a drive $\eta(\tau)=-\eta_0\cos^2(\omega_d \tau)$ which is periodic with frequency $2\omega_d$, the evolution operator of the cavity mode $(\hat{a}^{\dagger}, \hat{a})$ is decomposed using Floquet's theorem as: 
\begin{equation} \label{eq:FloquetTheorem}
\mathbf{U}(\tau) = \mathbf{G}(\tau) e^{-\mathrm{i} \mathbf{\mu} \tau },
\end{equation}
where  $\mathbf{G}(\tau)$ and $\mathbf{\mu}$ are both $2\times2$ matrices acting on the spinor space of the cavity mode $(\hat{a}^{\dagger}, \hat{a})$. $\mathbf{G}(\tau)$ is periodic with the same period as the drive and describes the periodic micro-motion of the system while the Floquet exponent matrix $e^{-\mathrm{i} \mathbf{\mu} \tau} $ describes its long-term behavior.\\
In the supplementary material, it is shown that the reflection coefficient can be decomposed in Fourier series as:
\begin{equation}
\tilde{r}(\tau,\omega) = \tilde{r}_b (\omega) + \sum_n \tilde{r}_n(\omega) e^{\mathrm{i} 2 n \omega_d \tau}
\end{equation}  where $\tilde{r}_b (\omega) $ corresponds to a non-resonant background contribution to the reflectivity and the $n^{th}$ order Fourier harmonic $\tilde{r}_n (\omega)$ can be expressed as:
\begin{equation}
    \frac{\tilde{r}_n(\omega)}{- \mathrm{i} \gamma } =  ~ \frac{\tilde{c}_n(\omega - \tilde{\omega}_d)+\tilde{d}_n}{(\omega - \tilde{\mu}_{+})(\omega - \tilde{\mu}_{-})}
    \label{fitting_formula}
\end{equation}
Here, $\gamma$ corresponds to the radiative power decay rate, the Floquet eigenvalues $\tilde{\mu}_{\pm}$ are defined through $\tilde{\mu}_{\pm}=\tilde{\omega}_d\pm\Delta \tilde{\mu}$ with $\tilde{\omega}_d\overset{\text{def}}{=}\omega_d-i\Gamma$ and can be obtained through diagonalization of the Floquet matrix $\mathbf{\mu}$, while the coefficients $(\tilde{c}_n, \tilde{d}_n)$ are related to the Fourier coefficients of the $\mathbf{G}(\tau)$ matrix. Their complete expression can be found in the supplementary material.

\subsubsection*{Spectral lineshapes (Figure 3a,d)}
The experimental lineshapes of $\tilde{r}_0(\omega)$ and $\tilde{r}_{\pm1}(\omega)$ are obtained from a short term Fourier transform of the time domain data (Fig. 2b) performed over a sliding window $[\tau,\tau+2\pi/\omega_d]$ as a function of time $\tau$. They are presented in Figure 3a for $\tau\approx 24ps$. The complete temporal evolution of $\tilde{r}_0(\omega)$ and $\tilde{r}_{\pm1}(\omega)$ is reported in the Supplementary material.
Theoretically, they are obtained from numerical computation of formula (\ref{fitting_formula}). \\
\subsubsection*{Experimental determination of the Floquet eigenvalues (Figure 4a)}
The Floquet eigenvalues presented in Fig. 4a are obtained from fitting the experimental lineshapes according to equation (\ref{fitting_formula}). Other coefficients are reported in the supplementary material. The fit is simultaneously performed on the real and imaginary parts of $\tilde{r}_0(\omega)$ and $\tilde{r}_{-1}(\omega)$ together.\\
As shown in the supplementary material, the fitting procedure is benchmarked on theoretical data to ensure that it accurately reproduces the eigenvalues that can obtained independently from diagonalization of the Floquet matrix. 

\subsubsection*{Stability diagram (Figure 4b)}
 In the absence of dissipation, the imaginary part of the eigenvalues $\mu_{\pm}$ determine wether the system is stable ($\mathrm{Im}(\mu_{\pm})<0$) or unstable ($\mathrm{Im}(\mu_{\pm})>0$), as quantities evolve as $e^{\mathrm{Im}(\mu_{\pm})\tau} $ at long times.  In the presence of dissipation characterized by a total decay rate $\Gamma= \frac{\gamma + \gamma^{\prime} }{2}$, it can be shown that eigenvalues of $\mathbf{\mu}$ undergo a shift by $-i \Gamma$ while $\mathbf{G}(t)$ remains unaffected. This allows to conclude directly on the stability of the dynamics from a contour line defined by $\frac{|\Delta \Gamma|}{\omega_\mathrm{c}^0}=\frac{\Gamma}{\omega_\mathrm{c}^0}$ and which marks the transition between stable ($\frac{|\Delta \Gamma|}{\omega_\mathrm{c}^0}<\frac{\Gamma}{\omega_\mathrm{c}^0}$) and unstable dynamics ($\frac{|\Delta \Gamma|}{\omega_\mathrm{c}^0}>\frac{\Gamma}{\omega_\mathrm{c}^0}$).
The total normalized decay rate $\frac{\Gamma}{\omega_\mathrm{c}^0}=0.12$ presented in Fig. 4b. corresponds to the experimentally measured one and can be found in Table 1.

\bibliographystyle{unsrt}
\bibliography{Main.bib}

\end{document}